\documentclass[apjl]{emulateapj}

\usepackage{graphicx}
\usepackage{wasysym}
\usepackage{amsbsy}
\usepackage{natbib}
\usepackage[colorlinks,urlcolor=cyan,citecolor=blue,linkcolor=blue]{hyperref} 

\shortauthors{} 
\shorttitle{}
\begin{document}

\title{The Albedos of {\it Kepler's} Close-in super-Earths}

\author{
Brice-Olivier Demory\altaffilmark{1,2}
}

\altaffiltext{1}{Astrophysics Group, Cavendish Laboratory, J.J. Thomson Avenue, Cambridge CB3 0HE, UK. bod21@cam.ac.uk}
\altaffiltext{2}{Department of Earth, Atmospheric and Planetary Sciences, Massachusetts Institute of Technology, 77 Massachusetts Ave., Cambridge, MA 02139, USA.}

\begin{abstract}
Exoplanet research focusing on the characterization of super-Earths is currently limited to those handful targets orbiting bright stars that are amenable to detailed study. This Letter proposes to look at alternative avenues to probe the surface and atmospheric properties of this category of planets, known to be ubiquitous in our galaxy. I conduct Markov Chain Monte Carlo lightcurve analyses for 97 {\it Kepler} close-in $R_P \lesssim 2.0 R_{\oplus}$ super-Earth candidates with the aim to detect their occultations at visible wavelengths. Brightness temperatures and geometric albedos in the {\it Kepler} bandpass are constrained for 27 super-Earth candidates. A hierarchical Bayesian modeling approach is then employed to characterize the population-level reflective properties of these close-in super-Earths. I find median geometric albedos $A_g$ in the {\it Kepler} bandpass ranging between 0.16 and 0.30, once decontaminated from thermal emission. These super-Earths geometric albedos are statistically larger than for hot Jupiters, which have medians $A_g$ ranging between 0.06 and 0.11. A subset of objects, including Kepler-10b, exhibit significantly larger albedos ($A_g\gtrsim$0.4). I argue that a better understanding of the incidence of stellar irradiation on planetary surface and atmospheric processes is key to explain the diversity in albedos observed for close-in super-Earths.
\end{abstract}


\section{Introduction}

Super-Earth mass and radius measurements leave significant degeneracy regarding their bulk composition. Even with an excellent precision on a super-Earth's physical parameters, similar masses and radii could be interpreted very differently \citep{Rogers:2010a,Miller-Ricci:2010}. One workaround is to have recourse to other types of measurements, such as transmission spectroscopy, occultation and phase-curve photometry to constrain the atmospheric and possibly surface properties. 
All these techniques, requiring bright ($K<9$) host stars, have been successfully applied to several hot-Jupiters systems and boosted exoplanet characterization to a level far beyond the mass-radius interpretation alone \citep[e.g.,][]{Deming:2009b}. To date however, only three transiting super-earths orbiting bright stars have been detected: GJ1214b \citep{Charbonneau:2009}, 55Cnc\,e \citep{Demory:2011,Winn:2011a} and more recently, HD97658b \citep{Dragomir:2013a}. All these super-Earths have mean densities favoring volatile-rich compositions.

Remarkably, {\it Kepler} has revealed a large population of smaller, close-in hot super-Earths similar to CoRoT-7b \citep{Leger:2009} and Kepler-10b \citep{Batalha:2011}. These strongly irradiated rocky planets could potentially harbor minimal atmospheres resulting from ground sublimation. Such atmospheres would be primarily made of monoatomic Na and O, O$_2$ and SiO \citep{Schaefer:2009,Miguel:2011}. Close-in super-Earths are expected to be tidally locked to their host stars, resulting in large temperature contrasts between the planet's day side and night side, to the point where the atmosphere would condense out at the day-night terminator \citep{Castan:2011,Heng:2012c}.

In this Letter, I propose an approach to explore the reflective properties of close-in super-Earths with no dependence on theoretical models. I conduct a search for occultations over a large sample of super-Earths so as to identify common patterns of their surface and/or atmosphere properties. 

This Letter is organized as follows. Section 2 describes the photometric analysis, including data reduction and light-curve modeling. Section 3 presents the hierarchical Bayesian framework used to interpret the results while Section 4 discusses the possible origin of visible flux from {\it Kepler}'s close-in super-Earths.

\section{Photometric Analysis} 
\label{analysis}

\subsection{Selection of Candidates}

The target selection is based on the {\it Kepler} quarters Q0 through Q13 \citep[see][for Q1-Q8]{Burke:2014}, which represents 3 years of quasi-continuous monitoring obtained between May 2009 and June 2012. All {\it Kepler} Objects of Interest (KOI) with radii $R_{P}<2.25R_{\oplus}$ are kept. Since this study focuses on how the incident radiation is processed by the planet surface/atmosphere, only those KOI that receive significant irradiation with orbital periods $P < 10$ days are retained. These two selection criteria result in 97 KOIs that are not flagged as false positives on MAST.

\subsection{Light-curve Modelling}

The Q0-Q13 long-cadence simple aperture photometry \citep{Jenkins:2010a} is retrieved from MAST\footnote{http://archive.stsci.edu/kepler/} for each of these 97 planet candidates. Instrumental systematics are mitigated by fitting the first four cotrending basis vectors (CBV) to each quarter using the PyKE software \citep{Still:2012}. For each quarter, the degree of photometric dilution is estimated by using the contamination factor in the fits file headers. Each quarter is then normalized to the median.

The goal of this analysis is to refine the transit parameters and to characterize the occultation in the {\it Kepler} bandpass for each planet candidate. For this purpose, I use the Markov Chain Monte Carlo (MCMC) implementation presented in \citet{Gillon:2012a}. The long-cadence 29.9-min exposure time is taken into account by resampling the photometric time-series to 1-min cadence directly in the MCMC framework \citep[e.g.,][]{Kipping:2010c}. 

I assume a quadratic law for the limb-darkening (LD) and use $c_1=2u_1+u_2$ and $c_2=u_1-2u_2$ as jump parameters, where $u_1$ and $u_2$ are the quadratic coefficients. $u_1$ and $u_2$ are drawn from the theoretical tables of \citet{Claret:2011} for the corresponding $T_{eff}$ and log $g$ values extracted from the Q1-Q16 star properties catalog of \citet{Huber:2014}. Constraining the stellar density from the transit photometry is more difficult for super-Earths than for hot Jupiters \citep[e.g.,][]{Demory:2011a} because of the smaller planet/star radius ratio. I thus include Gaussian priors in the MCMC fit for the stellar radius, $T_{eff}$ and log $g$ values extracted from \citet{Huber:2014}.

Each MCMC fit has the following set of jump parameters: the planet/star radius ratio $R_{P}/R_{\star}$, the impact parameter $b$, the transit duration from first to fourth contact $T_{14}$, the time of minimum light $T_0$, the orbital period $P$, the occultation depth $dF_{occ}$, the two LD combinations $c_1$ and $c_2$ and the two parameters $\sqrt{e}\cos\omega$ and $\sqrt{e}\sin\omega$. I use Gaussian priors for $c_1$ and $c_2$, based on the theoretical tables. To improve the MCMC iteration efficiency and because the planet candidates have short orbital periods, I further impose Gaussian priors on $\sqrt{e}\cos\omega$ and $\sqrt{e}\sin\omega$ by centering the distributions on zero and assuming a standard deviation of 0.45 for both parameters. These priors prevents the MCMC fit from exploring high eccentricity configurations that seem highly unlikely for such systems \citep{Hadden:2013}. Negative occultation values are allowed in the MCMC fit to avoid biasing occultation posteriors toward positive values.

I divide the photometric time-series in $\sim$24 to 48~hr segments and fit for each of them the smooth photometric variations due to stellar variability with a time-dependent quadratic polynomial in the MCMC fit. Baseline model coefficients are determined at each step of the MCMC procedure for each lightcurve using a singular value decomposition method. The resulting coefficients are then used to correct the raw photometric lightcurves. For each data segment, correlated noise is accounted for following \citet{Gillon:2010a} to ensure reliable error bars on the fitted parameters. 

One MCMC fit consisting of two Markov chains of 100,000 steps is performed for each planet candidate and their convergence is checked using the Gelman-Rubin statistic criterion \citep{Gelman:1992}. MCMC fit results for all KOI are shown in Table~\ref{table}.

\subsection{Albedos, Brightness and Equilibrium temperatures}

The purpose of the present study is to characterize the planetary total emission in the {\it Kepler} bandpass, which is likely to have both thermal and reflected light components. Thus, the occultation in the {\it Kepler} bandpass $dF_{occ, kep}$ alone does not unambiguously provide a direct estimate of the geometric albedo nor the planet's temperature. I therefore define in the following the ``total'' albedo, as being a direct translation of the full occultation depth into an albedo estimate.

Both the total albedo and brightness temperature are useful means to convert the occultation depth into physical quantities. I use the posterior distributions functions obtained from the MCMC analyses to compute total albedo and brightness temperature values for all planet candidates. 

The total albedo in the {\it Kepler} bandpass is:

\begin{equation}
A_t=dF_{occ,kep}\left(\frac{a}{Rp}\right)^2
\end{equation}

Assuming a planetary blackbody spectrum, the planetary brightness temperature $T_B$ in the {\it Kepler} bandpass is obtained from the following equation: 

\begin{equation}
dF_{occ,kep}=\frac{R_{P}^2}{R_{\star}^2}\frac{\int B_{\lambda}(T_B)\Gamma_{\lambda}d\lambda}{\int F^{\star}_{\lambda}\Gamma_{\lambda}d\lambda}
\end{equation}

where $\Gamma_{\lambda}$ is the {\it Kepler} transmission function and $B_{\lambda}$ the Planck function. The stellar flux density $F^{\star}_{\lambda}$ is obtained by matching each host's properties \citep{Huber:2014} to the closest MARCS stellar model of \citet{Gustafsson:2008}. 

Several planet candidates that are part of the sample exhibit a moderate photometric signal-to-noise ratio and a short transit duration that, combined to long-cadence time-series, prevent from obtaining a precise estimate of $\frac{a}{R_{\star}}$ \citep[see also][]{Sanchis-Ojeda:2013}. As a consequence, a mediocre precision on both the albedo and equilibrium temperature is derived for these candidates. In a first step, I therefore base the analysis on the brightness temperature instead of the albedo.

\subsection{False Positive Assessment and Occultation Detectability}

\begin{figure*}
\centering
   \epsscale{1.0}\plottwo{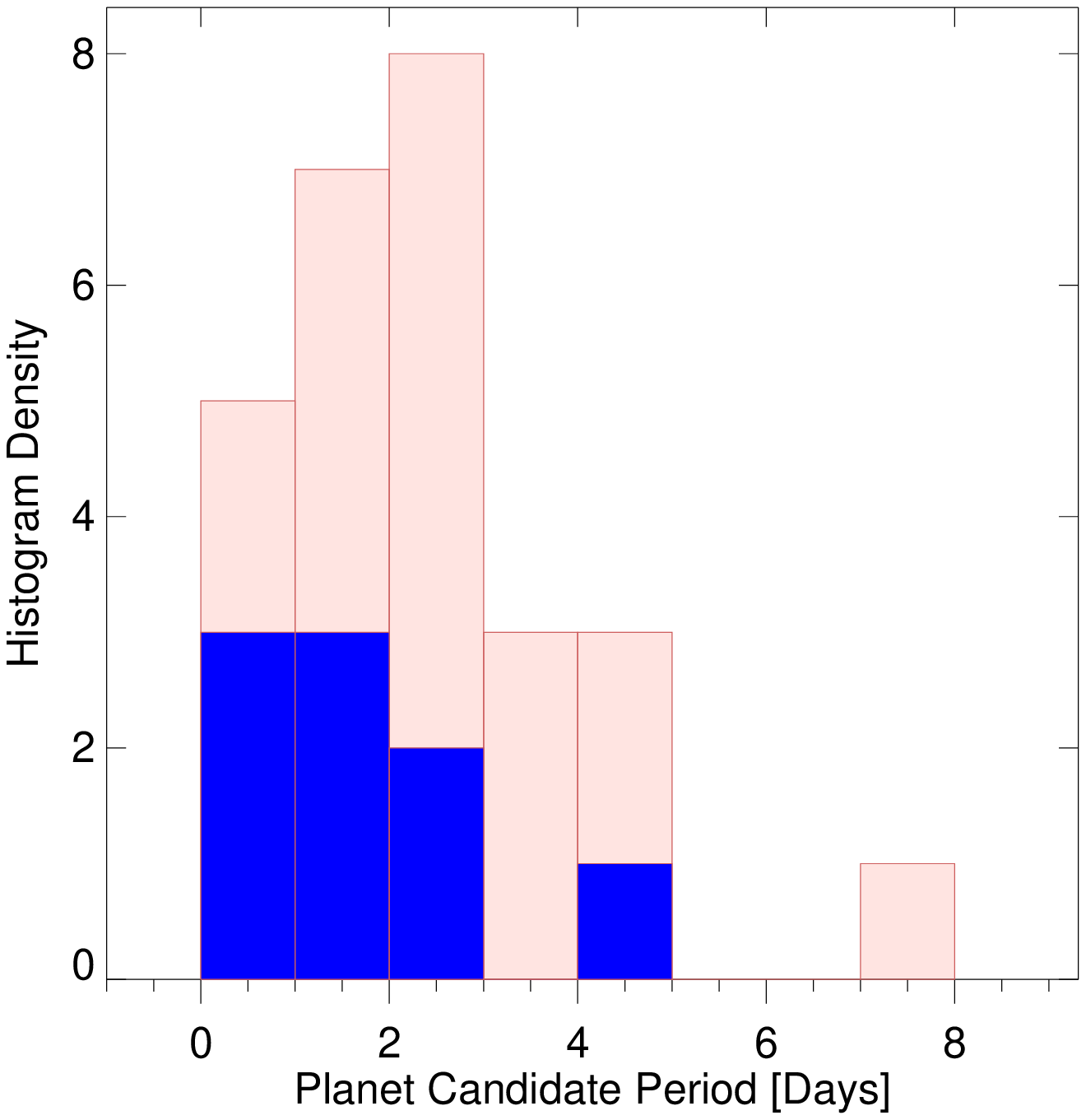}{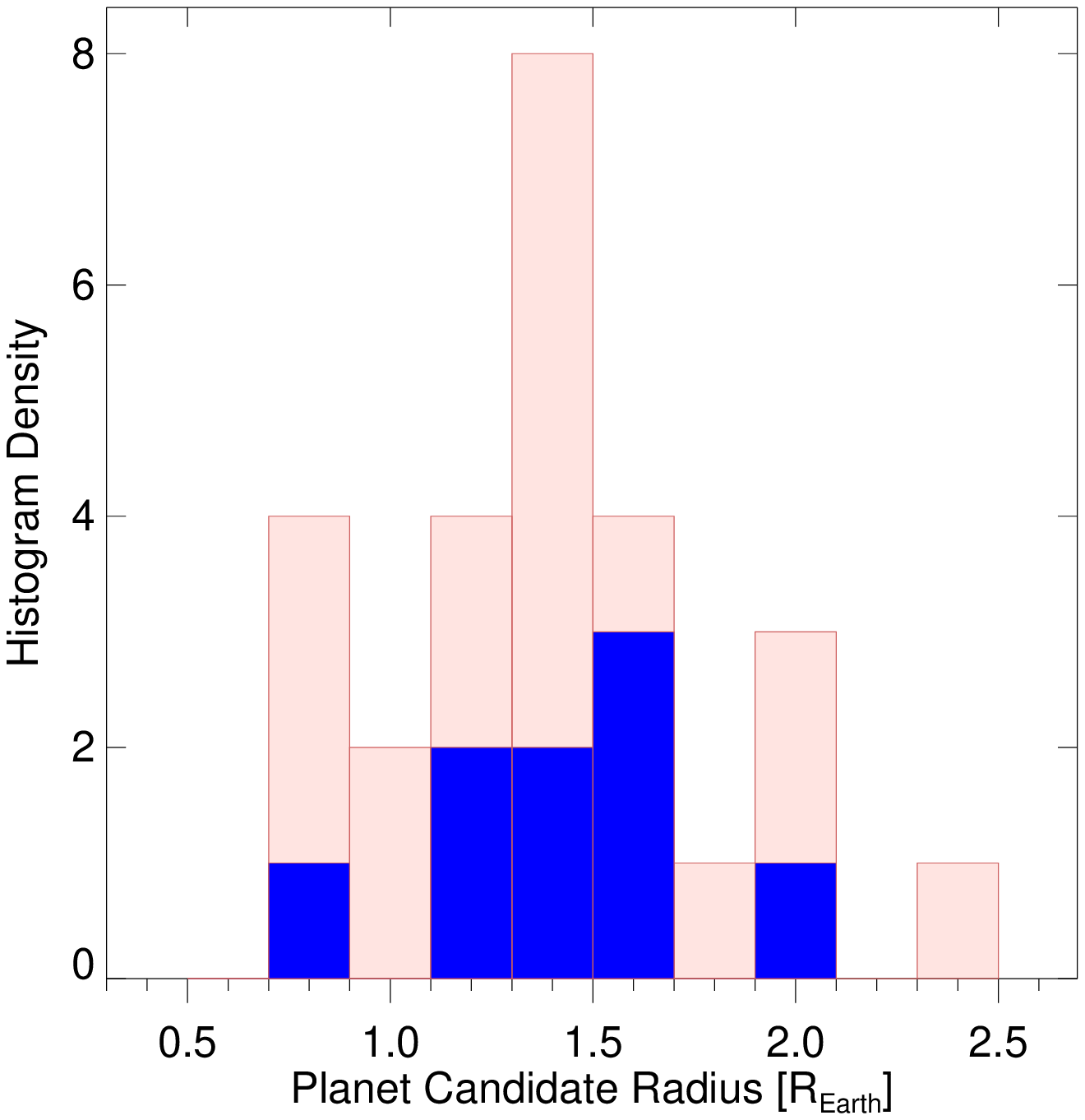}
  \caption{{\bf Properties of the selected {\it Kepler} super-Earth size candidates.} {\it Left:} Histogram of orbital periods. {\it Right:} Histogram of radii. The red bars show the 18 KOIs for which the occultation depth is compatible with 0 at the 1-$\sigma$ level. The blue bars indicate the 9 objects that have a determination of their brightness temperature.}  \label{fig_prop}
\end{figure*}

False positives are likely to contaminate the sample. In the radius range considered in this study, it is expected that the corresponding false positive rate ranges from 5 to 15\% \citep{Fressin:2013}. All obvious false-positives have been withdrawn during the Q1-Q8 \citep{Burke:2014} extensive vetting effort. I choose $A_t$ as a means to discard false positives \citep[e.g.,][]{Batalha:2010,Demory:2011a}. I keep only those planet candidates that have a total albedo uncertainty less than 1.0. This criterion results in a list of 27 candidates, representing 28\% of the initial sample. Keeping candidates with $\sigma_{A_t} < 1.0$ is a conservative approach that eliminates most high-albedo false positives but also low-SNR planet candidates for which no occultation is detected. This selection criterion does not bias the results about non detections, as high-SNR planetary candidates for which no occultation is detected (i.e., low albedo) will have precise uncertainties on their albedos and will therefore be retained. I use a $T_B$ vs. period distribution as a means to identify remaining false positives. I find that KOI 2272.01 \citep[see also][]{Ofir:2013}, 2545.01 and 2636.01 are likely diluted eclipsing binaries. Among the 27 KOIs resulting from this geometric albedo based selection, 4 only have periods longer than 4 days and all have total albedos $< 1$. Radius and orbital period distributions for the selected candidates are shown on Figure~\ref{fig_prop}. As expected, I find that the number of candidates showing an occultation decreases with increasing orbital period, strengthening the planetary nature of the targets \citep{Slawson:2011}. 

\subsection{Individual super-Earth Brightness temperatures}

\begin{figure}
\centering
   \epsscale{1.0}\plotone{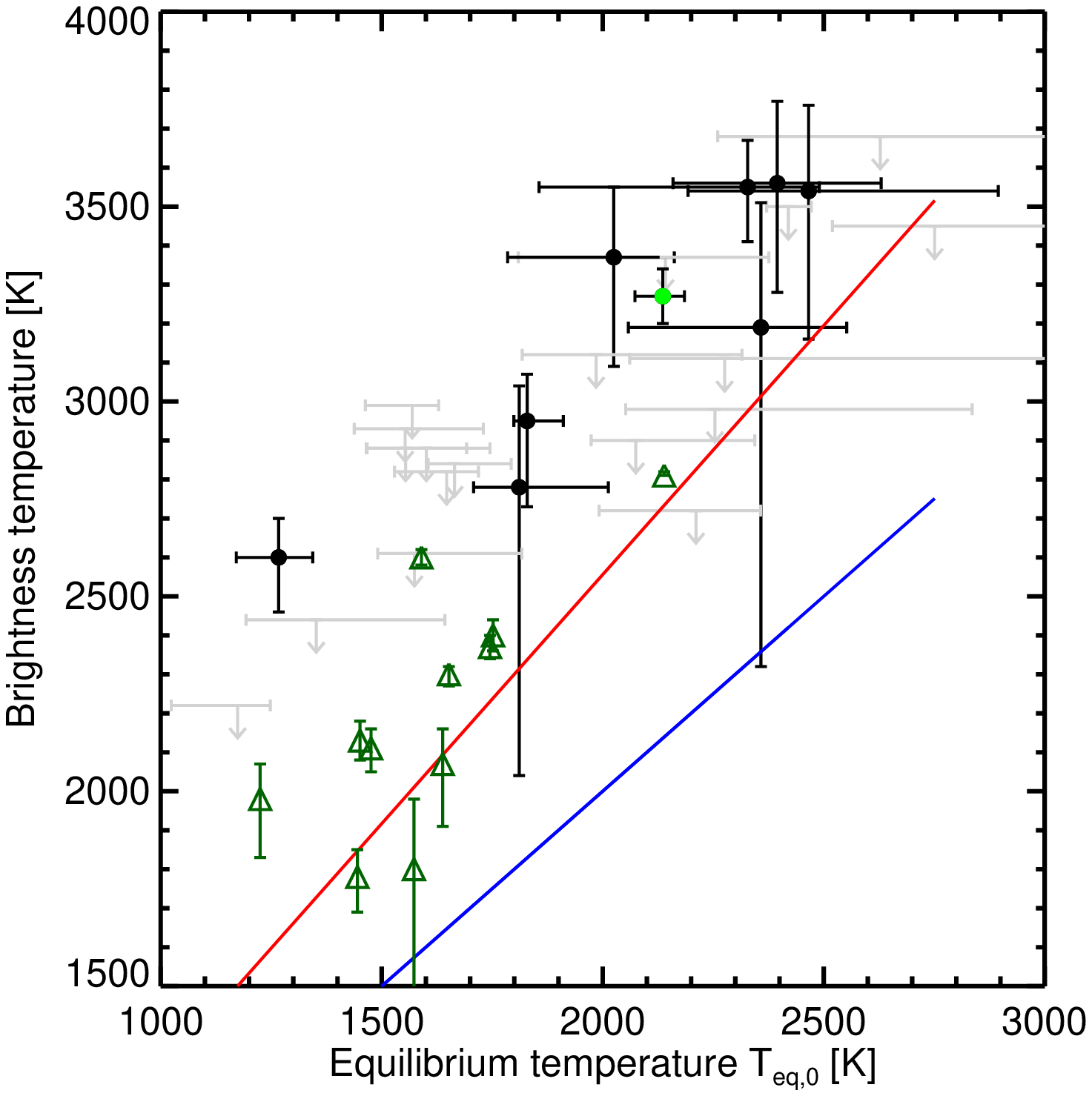}
  \caption{{\bf Brightness temperatures of {\it Kepler} super-Earth-sized candidates.} The brightness temperatures measured from the occultation in the {\it Kepler} bandpass are shown for the 27 selected super-Earth candidates. The red and blue lines depict the equilibrium temperature, assuming null and efficient heat redistribution respectively, both for a zero Bond Albedo. The green disk shows Kepler-10b and the triangles are the {\it Kepler} published hot Jupiters \citep{Heng:2013a} that have a constraint on their occultation depth. Planet candidates for which the occultation depth is compatible with 0 at the 1-$\sigma$ level are shown in gray as 1-$\sigma$ upper limits.}  \label{fig_teq_tb}
\end{figure}

Figure~\ref{fig_teq_tb} shows for each planet candidate the brightness temperature derived from the occultation depth in the {\it Kepler} bandpass vs. the planet's equilibrium temperature $T_{eq,0} = T_{\star} \sqrt{\frac{R_{\star}}{2a}}$, assuming a null Bond albedo and an efficient heat recirculation \citep{Hansen:2008}. The red line is the maximum equilibrium temperature $T_{eq,{\rm max}}$ with no heat recirculation while the blue line is for an efficient redistribution, both with a null albedo. The planet candidates that are part of the sample span equilibrium temperatures ranging from 1200K to 2800K.

\begin{figure}
\centering
   \epsscale{1.0}\plotone{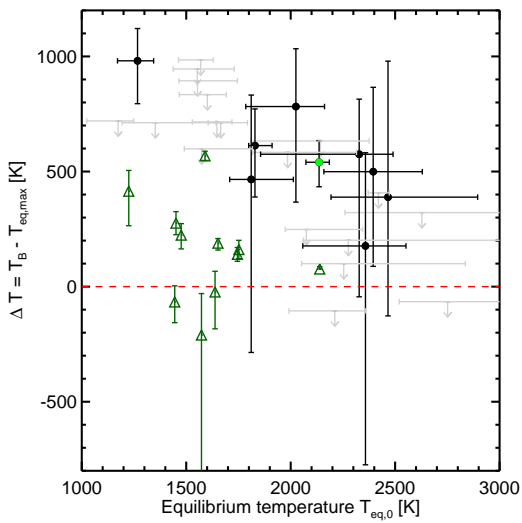}
  \caption{{\bf Brightness temperature excesses of {\it Kepler} super-Earth-sized candidates.} The brightness temperature excesses ($\Delta T=T_{B}-T_{eq,max}$) are shown for the 27 selected planetary candidates as a function of the equilibrium temperature $T_{eq,0}$. Triangles depict {\it Kepler} published hot Jupiters \citep{Heng:2013a} that have a constraint on their occultation depth. The green disk represents Kepler-10b. Planet candidates for which the occultation depth is compatible with 0 at the 1-$\sigma$ level are shown in gray as 1-$\sigma$ upper limits.}  \label{fig_teq_tdiff}
\end{figure}

To investigate more precisely how $T_{B}$ evolves with the equilibrium temperature, I compute $\Delta T = T_{B} - T_{eq,{\rm max}}$ for each super-Earth candidate on Figure~\ref{fig_teq_tdiff}. The red dash line depicts planets with $T_B = T_{eq,{\rm max}}$. It is important to note that across the wide range of equilibrium temperatures studied here, the sensitivity to small $\Delta T$ is better in the elevated $T_{eq}$ regime than in the lower end. Upper limits on $\Delta T$ are obtained for planets in the less populated area of low $T_{eq}$/low $\Delta T$ but no objects in the high $T_{eq}$/high $\Delta T$ are found, where the sensitivity is optimal. I pursue the analysis in a Bayesian framework to explore the incidence of this bias on the results. I find no correlation between the occultation's SNR and $\Delta T$ (Spearman's rank correlation coefficient of 0.25).

\begin{deluxetable*}{lllllllllll}
\tablecaption{System Parameters, Brightness Temperatures and Albedos of this study's super-Earth sample. \label{table}}
%\rotate
\tabletypesize{\scriptsize}
\startdata

KOI & Period & $R_P$             & Rp/R$_{\star}$ & a/R$_{\star}$ & $T_{eq,0}$ & Occ. depth & $T_B$ & $A_t$ & $A_{g,max}$ & $A_{g,min}$  \\
       & [days]  & [$R_{\oplus}$] &                &     &  [K] &     [ppm]    &   [K]       &   &           &                         \\
\hline

  69.01   & 4.727   & $ 1.51_{  -0.05}^{+ 0.07}$ & $0.01564_{  -0.00042}^{   +0.00060}$ & $   9.998_{-1.500}^{+ 1.200}$ & $ 1267_{  -96}^{+   77}$ & $ 2.2_{-0.8}^{+ 0.7}$ & $   2600_{ -140}^{+  100}$ & $   0.88_{-0.34}^{+ 0.48}$ &  0.88&    0.87 \\
  70.02   & 3.696   & $ 1.87_{  -0.21}^{+ 0.35}$ & $0.01930_{  -0.00160}^{   +0.00120}$ & $   8.167_{-1.900}^{+ 3.500}$ & $ 1352_{ -159}^{+  291}$ & $-1.6_{-2.4}^{+ 2.4}$ & $    <2440$ & $   <0.43$ &  $<0.43$&    $<0.41$ \\
  72.01   & 0.837   & $ 1.39_{  -0.03}^{+ 0.04}$ & $0.01258_{  -0.00012}^{   +0.00020}$ & $   3.468_{-0.200}^{+ 0.150}$ & $ 2136_{  -63}^{+   49}$ & $ 7.4_{-1.0}^{+ 1.1}$ & $   3270_{  -70}^{+   70}$ & $   0.56_{-0.10}^{+ 0.09}$ &  0.54&    0.41 \\
  85.02   & 2.155   & $ 1.43_{  -0.05}^{+ 0.08}$ & $0.00956_{  -0.00026}^{   +0.00048}$ & $   3.891_{-0.770}^{+ 0.510}$ & $ 2211_{ -219}^{+  146}$ & $-0.5_{-0.9}^{+ 0.8}$ & $    <2720$ & $   <0.13$ &  $<0.11$&   $<-0.05$ \\
 167.01   & 4.920   & $ 2.39_{  -0.29}^{+ 1.22}$ & $0.01953_{  -0.00025}^{   +0.00040}$ & $   8.559_{-1.000}^{+ 0.530}$ & $ 1569_{ -106}^{+   60}$ & $ 2.1_{-3.8}^{+ 3.7}$ & $    <2990$ & $   <1.11$ &  $<1.11$&   $ <1.07$ \\
 262.01   & 7.813   & $ 2.04_{  -0.05}^{+ 0.05}$ & $0.01225_{  -0.00015}^{   +0.00014}$ & $   3.306_{-0.110}^{+ 0.120}$ & $ 2420_{  -49}^{+   52}$ & $-2.6_{-3.8}^{+ 7.7}$ & $    <3500$ & $   <0.56$ &  $<0.52$&    $<0.32$ \\
 292.01   & 2.587   & $ 1.43_{  -0.15}^{+ 0.28}$ & $0.01414_{  -0.00064}^{   +0.00035}$ & $   6.967_{-0.760}^{+ 1.700}$ & $ 1554_{  -89}^{+  191}$ & $ 1.1_{-2.8}^{+ 2.8}$ & $    <2780$ & $   <0.95$ &  $<0.94$&    $<0.91$ \\
 299.01   & 1.542   & $ 1.48_{  -0.06}^{+ 0.20}$ & $0.01592_{  -0.00030}^{   +0.00056}$ & $   5.733_{-0.790}^{+ 0.460}$ & $ 1647_{ -118}^{+   72}$ & $ 2.3_{-2.8}^{+ 2.8}$ & $    <2820$ & $   <0.66$ &  $<0.65$&   $ <0.60$ \\
 321.01   & 2.426   & $ 1.31_{  -0.08}^{+ 0.23}$ & $0.01280_{  -0.00043}^{   +0.00022}$ & $   5.847_{-0.350}^{+ 0.870}$ & $ 1665_{  -60}^{+  128}$ & $-0.6_{-1.9}^{+ 1.9}$ & $    <2700$ & $   <0.40$ &  $<0.39$&   $ <0.34$ \\
 343.02   & 2.024   & $ 1.92_{  -0.29}^{+ 0.29}$ & $0.01471_{  -0.00065}^{   +0.00040}$ & $   4.996_{-0.540}^{+ 1.100}$ & $ 1811_{ -103}^{+  202}$ & $ 3.2_{-3.0}^{+ 2.9}$ & $   2780_{ -740}^{+  260}$ & $   0.37_{-0.34}^{+ 0.41}$ &  0.36&    0.29 \\
 665.02   & 1.612   & $ 1.12_{  -0.14}^{+ 0.58}$ & $0.00962_{  -0.00040}^{   +0.00045}$ & $   3.253_{-0.610}^{+ 0.620}$ & $ 2395_{ -236}^{+  235}$ & $ 6.2_{-2.5}^{+ 2.6}$ & $   3560_{ -280}^{+  210}$ & $   0.71_{-0.36}^{+ 0.52}$ &  0.67&    0.45 \\
 975.01   & 2.786   & $ 1.53_{  -0.03}^{+ 0.02}$ & $0.00784_{  -0.00011}^{   +0.00007}$ & $   5.617_{-0.170}^{+ 0.500}$ & $ 1829_{  -30}^{+   82}$ & $ 0.9_{-0.4}^{+ 0.3}$ & $   2950_{ -220}^{+  120}$ & $   0.51_{-0.20}^{+ 0.20}$ &  0.51&    0.43 \\
1128.01   & 0.975   & $ 1.21_{  -0.09}^{+ 0.18}$ & $0.01386_{  -0.00087}^{   +0.00029}$ & $   3.495_{-0.310}^{+ 0.900}$ & $ 2075_{ -101}^{+  269}$ & $ 2.2_{-3.0}^{+ 2.6}$ & $    <2900$ & $   <0.30$ &  $<0.28$&    $<0.15$ \\
1169.01   & 0.689   & $ 1.26_{  -0.06}^{+ 0.30}$ & $0.01304_{  -0.00044}^{   +0.00230}$ & $   2.970_{-1.200}^{+ 0.400}$ & $ 2328_{ -472}^{+  162}$ & $13.5_{-3.1}^{+ 3.0}$ & $   3550_{ -140}^{+  120}$ & $   0.70_{-0.48}^{+ 0.35}$ &  0.67&    0.48 \\
1824.02   & 1.678   & $ 1.67_{  -0.37}^{+ 1.01}$ & $0.01156_{  -0.00027}^{   +0.00064}$ & $   4.804_{-1.100}^{+ 0.610}$ & $ 2025_{ -240}^{+  137}$ & $ 5.3_{-2.3}^{+ 2.2}$ & $   3370_{ -280}^{+  180}$ & $   0.92_{-0.40}^{+ 0.58}$ &  0.90&    0.78 \\
1883.01   & 2.707   & $ 1.00_{  -0.16}^{+ 0.48}$ & $0.00844_{  -0.00069}^{   +0.00022}$ & $   3.888_{-0.660}^{+ 2.000}$ & $ 2254_{ -202}^{+  582}$ & $-0.7_{-1.2}^{+ 1.3}$ & $    <2980$ & $   <0.28$ &  $<0.25$&    $<0.06$ \\
1890.01   & 4.336   & $ 1.58_{  -0.05}^{+ 0.08}$ & $0.00970_{  -0.00021}^{   +0.00039}$ & $   7.248_{-1.200}^{+ 0.810}$ & $ 1601_{ -134}^{+   91}$ & $-0.2_{-1.4}^{+ 1.5}$ & $    <2880$ & $   <0.84$ &  $<0.83$&    $<0.79$ \\
1929.02   & 3.293   & $ 1.37_{  -0.30}^{+ 0.22}$ & $0.00921_{  -0.00110}^{   +0.00034}$ & $   3.117_{-0.580}^{+ 2.000}$ & $ 2276_{ -215}^{+  731}$ & $ 0.5_{-1.9}^{+ 2.3}$ & $    <3110$ & $   <0.33$ &  $<0.30$&    $<0.12$ \\
1937.01   & 1.411   & $ 1.21_{  -0.14}^{+ 0.14}$ & $0.01842_{  -0.00066}^{   +0.00150}$ & $   6.863_{-1.700}^{+ 0.770}$ & $ 1174_{ -150}^{+   74}$ & $-0.4_{-4.0}^{+ 3.8}$ & $    <2220$ & $   <0.53$ &  $<0.53$&    $<0.52$ \\
1961.01   & 1.908   & $ 1.94_{  -0.82}^{+ 1.19}$ & $0.01078_{  -0.00051}^{   +0.00039}$ & $   6.352_{-0.890}^{+ 1.400}$ & $ 1553_{ -115}^{+  177}$ & $ 0.8_{-2.6}^{+ 2.3}$ & $    <2930$ & $   <1.08$ &  $<1.08$&    $<1.04$ \\
1964.01   & 2.229   & $ 0.74_{  -0.10}^{+ 0.13}$ & $0.00710_{  -0.00047}^{   +0.00013}$ & $   6.205_{-0.630}^{+ 1.900}$ & $ 1574_{  -83}^{+  243}$ & $-0.1_{-0.7}^{+ 0.6}$ & $    <2610$ & $   <0.43$ &  $<0.42$&    $<0.39$ \\
2049.01   & 1.569   & $ 1.40_{  -0.16}^{+ 0.92}$ & $0.01174_{  -0.00039}^{   +0.00073}$ & $   3.617_{-0.880}^{+ 0.570}$ & $ 2358_{ -300}^{+  194}$ & $ 3.8_{-3.5}^{+ 3.4}$ & $   3190_{ -870}^{+  320}$ & $   0.36_{-0.27}^{+ 0.43}$ &  0.33&    0.12 \\
2072.01   & 1.543   & $ 0.91_{  -0.12}^{+ 0.40}$ & $0.01038_{  -0.00120}^{   +0.00027}$ & $   2.442_{-0.390}^{+ 1.600}$ & $ 2751_{ -231}^{+  903}$ & $ 1.8_{-5.2}^{+ 4.4}$ & $    <3450$ & $   <0.34$ &  $<0.27$&   $<-0.05$ \\
2079.01   & 0.694   & $ 0.83_{  -0.16}^{+ 0.31}$ & $0.00681_{  -0.00059}^{   +0.00049}$ & $   2.560_{-0.550}^{+ 0.880}$ & $ 2466_{ -273}^{+  429}$ & $ 4.4_{-2.3}^{+ 1.9}$ & $   3540_{ -380}^{+  220}$ & $   0.62_{-0.29}^{+ 0.86}$ &  0.58&    0.34 \\
2332.01   & 3.701   & $ 1.39_{  -0.75}^{+ 1.11}$ & $0.00795_{  -0.00055}^{   +0.00033}$ & $   3.682_{-0.580}^{+ 1.200}$ & $ 1985_{ -167}^{+  330}$ & $-0.8_{-3.6}^{+ 2.7}$ & $    <3120$ & $   <0.58$ &  $<0.56$&    $<0.46$ \\
2470.01   & 2.175   & $ 0.73_{  -0.07}^{+ 0.31}$ & $0.00795_{  -0.00051}^{   +0.00079}$ & $   3.584_{-1.100}^{+ 0.760}$ & $ 2142_{ -333}^{+  234}$ & $ 0.3_{-3.1}^{+ 3.3}$ & $    <3370$ & $   <0.73$ &  $<0.71$&    $<0.56$ \\
2492.01   & 0.985   & $ 0.90_{  -0.11}^{+ 0.35}$ & $0.00926_{  -0.00081}^{   +0.00061}$ & $   2.658_{-0.730}^{+ 1.100}$ & $ 2628_{ -368}^{+  547}$ & $ 2.1_{-5.4}^{+ 5.0}$ & $    <3680$ & $   <0.58$ &  $<0.53$&    $<0.25$ \\

\enddata
\tablecomments{Planet candidates for which the occultation depth is compatible with 0 have their $T_B$ and albedo values shown as 1-$\sigma$ upper limits.}
\end{deluxetable*}

\section{Constraining $T_B$ and $A_g$ at the Population-level using Hierarchical Bayesian Modeling} 

The data presented in Table~\ref{table} include several non-detections and upper limits. I retain them intentionally to avoid biasing the interpretations toward a sub-group of objects that would not be representative of the entire population of planets studied here. Furthermore, the maximum-likelihood estimates alone cannot thoroughly represent the probability intervals of  $T_B$ or $A_g$.
I choose an approach based on a hierarchical Bayesian model that follows the method described by \citet{Hogg:2010} and Rogers (submitted). In summary, I assume a parameterized model for the true distribution of $T_B$ and $A_g$ that is constrained at the population-level by the individual planet MCMC posterior probability distributions (PPD). I briefly describe the method and present the results below.

\subsection{Methodology}

I consider $N$ stars indexed by $n$. Each star is orbited by one or more planets whose transit and occultation parameters derived from the MCMC are denoted by $\boldsymbol{\theta_n}$. The {\it Kepler} photometric time-series obtained on each star are denoted by $\boldsymbol{y_n}$.
In Sect.~\ref{analysis} the MCMC procedure computes the PPD for each parameter in 
$\boldsymbol{\theta_n}$:

\begin{equation}
p(\boldsymbol{\theta_n} \mid \boldsymbol{y_n})=\frac{1}{Z_n}p(\boldsymbol{y_n} \mid \boldsymbol{\theta_n})p_0(\boldsymbol{\theta_n})
\end{equation}

where $Z_n$ is a normalization constant, $p(\boldsymbol{y_n} \mid \boldsymbol{\theta_n})$ is the likelihood and $p_0(\boldsymbol{\theta_n})$ is a prior chosen for each parameter in the MCMC fit. Uniform priors are chosen for all parameters but the LD coefficients and orbital eccentricity (Sect.~\ref{analysis}). From this point, individual planet properties need to be recast within a population-level joint posterior probability depending on the hyperparameters $\boldsymbol{\alpha}$:

\begin{equation}
p \left(\{ \boldsymbol{\theta_n} \},\boldsymbol{\alpha} \mid \{\boldsymbol{y_n}\}\right) \propto p \left(\{\boldsymbol{y_n}\} \mid \{\boldsymbol{\theta_n}\}\right) p\left(\{\boldsymbol{\theta_n}\} \mid \boldsymbol{\alpha} \right) p\left(\boldsymbol{\alpha}\right)
\label{eq2}
\end{equation}

This relies on the fact that the global PPD, likelihood and prior for all parameters are the product of individual PPD, likelihoods and priors respectively.

Since the goal is to constrain the population-level model parameters $\boldsymbol{\alpha}$, individual planetary parameters in equation~\ref{eq2} need to be marginalized to obtain the likelihood on 
$\boldsymbol{\alpha}$:

\begin{equation}
\mathcal{L_{\boldsymbol{\alpha}}}=\prod_{n=0}^N\int d\boldsymbol{\theta_n}p(\boldsymbol{y_n}\mid\boldsymbol{\theta_n})p(\boldsymbol{\theta_n}\mid\boldsymbol{\alpha})
\end{equation}

which can be approximated as:

\begin{equation}
\mathcal{L_{\boldsymbol{\alpha}}}\approx\prod_{n=0}^N\frac{1}{K}\sum_{k=0}^K\frac{p(\boldsymbol{\theta_{nk}}\mid\boldsymbol{\alpha})}{p_0(\boldsymbol{\theta_{nk}})}
\end{equation}

with

\begin{equation}
p(\boldsymbol{\theta_{nk}}\mid\boldsymbol{\alpha})\equiv\frac{f_{\boldsymbol{\alpha}}(\psi_{nk}) p_0(\boldsymbol{\theta_{nk}})}{p_0(\psi_{nk})}
\end{equation}

where $K$ is the number of samples in the PPD. This is a marginal likelihood where nuisance parameters are now under the form of priors. $\psi_{nk}$ is the population-level distribution that is searched for.

\subsection{Dependence of $T_B$ with the stellar incident flux}

I now tailor the general framework described above to investigate the dependence of the brightness temperature excess $\Delta T=T_{B}-T_{eq,{\rm max}}$ with equilibrium temperature $T_{eq,{\rm max}}$. I use a simple parameterized model to describe the intrinsic distribution of $\Delta T$ that depends on $\boldsymbol{\alpha}$:

\begin{equation}
f_{\boldsymbol{\alpha}}(\Delta T_{nk})=\gamma T_{eq,{\rm max,nk}}+\zeta
\label{eqf}
\end{equation}

with $\boldsymbol{\alpha}\equiv(\gamma,\zeta)$ and adopting uniform priors on $\gamma$ and $\zeta$. I use as inputs the $\Delta T$ PPD extracted from the MCMC by truncating the lower end of the $T_B$ distribution to $T_{eq,0}$. In other words, I hypothesize for each planet that $T_B$ cannot be below $T_{eq,0}$.

I find a median value $\gamma=-0.32^{+0.26}_{-0.24} $. Based on the current data, the $\Delta T$ vs. $T_{eq,0}$ trend is thus not robustly detected. As previously discussed, this could be due to the fact that a small $\Delta T$ is easier to characterize in the high-irradiation regime. The upper envelope on Fig.\ref{fig_teq_tdiff} represents a subset of super-Earths that exhibit large $\Delta T$ but that are not representative of the whole population of super-Earths studied here.

\subsection{Determination of a population-level $A_g$}

I use the same model as eq.~\ref{eqf} to infer the geometric albedo in the {\it Kepler} bandpass $A_g$.  The ``total albedo'' $A_t$ is distinguished from $A_{g,min}$ as well as $A_{g,max}$, which correspond to minimum and maximum albedo estimates after decontamination from $T_{eq,{\rm max}}$  and  $T_{eq,0}$ respectively \citep{Heng:2013a}. I find median $A_t = 0.32$, $A_{g,max} = 0.30$ (95\% upper limit of $0.56$) and $A_{g,min} = 0.16$ (95\% upper limit of $0.47$).

\section{Discussion} 
\label{discussion}

\subsection{Comparison With Published hot Jupiters}

I base this comparison sample on the hot Jupiters whose albedo and brightness temperature have been published in \citet{Heng:2013a}, shown as triangles on Figure~\ref{fig_teq_tb}. These planets are TrES-2b, Kepler-5b, Kepler-6b, Kepler-7b, Kepler-8b, Kepler-12b, Kepler-14b, Kepler-15b, Kepler-17b, Kepler-41b and HAT-P-7b. Most giant planets have brightness temperatures within $\sim$300K of their maximum equilibrium temperature. Super-Earth exoplanets discussed in this paper have larger scatter and exhibit brightness temperature excesses up to $\sim$1000K. Figure~\ref{fig_teq_tdiff} shows that hot Jupiters generally match a decreasing trend towards a null excess, as shown for HAT-P-7b \citep{Pal:2008}, which is the archetype of highly-irradiated, low-albedo hot Jupiters with a very low heat recirculation efficiency \citep{Christiansen:2010a,Spiegel:2010b}. 
There are also hot Jupiters having negligible $T_B$ excesses, owing to their low albedo such as TrES-2b \citep[e.g.,][]{Barclay:2012a}. Kepler-7b appears as an outlier on this figure because of its large albedo \citep{Demory:2013b}.

I find minimum and maximum geometric albedos for hot Jupiters of $A_{g,min}=0.06\pm0.08$ and $A_{g,max}=0.11\pm0.09$ respectively, which are statistically smaller than geometric albedos found for super-Earths.

\subsection{Close-in Super-Earth Properties}

Close-in super-Earths ($ 1.0<R_{P}<2.25R_{\oplus}$) have geometric albedos spanning a large range of values in the {\it Kepler} bandpass. I find that 21 out of the 27 candidates have radii $R_P\leq 1.6 R_{\oplus}$, suggesting that most of the targets are likely rocky \citep[][and Rogers, submitted]{Lopez:2013}.  For larger strongly irradiated planets such as 55Cnc\,e, one could not discard the possibility that a significant brightness temperature excess could be due to the fact that the {\it Kepler} bandpass probes the deep layers of an envelope of volatiles such as water \citep[see also][]{Cowan:2011a}. In such a case the incident heat recirculation across longitude at high pressure is more efficient, but a strong greenhouse effect would dramatically increase the brightness temperature. In the case of the close-in super-Earths that are part of this work, I argue that the source of reflected light may originate from 1) the surface, 2) Rayleigh scattering in an atmosphere that may have absorbers in the visible (Na and K) or 3) reflective clouds. 

\subsubsection{Atmospheres}
The range of geometric albedos that are observed for close-in super-Earths suggests that strong alkali metal absorption in the visible is not as generalized as it is for hot-Jupiters. The objects forming the upper envelope of brightness temperature excesses (Fig.~\ref{fig_teq_tdiff}) may provide information about the incidence of stellar irradiation on atmospheric visible absorption. Above $T_{eq,0}\sim2200$K, the observed brightness temperature can be explained by thermal emission alone. One possible explanation is that at these high temperatures, the column density of atmospheric constituents may be higher because of the increased evaporation rate of the planet surface \citep{Miguel:2011}. Such high-pressure atmospheres would have a higher concentration of alkali metals that would result in an increased absorption at visible wavelengths, hence a lower albedo. At cooler temperatures, the lower evaporation rate would yield an atmosphere with a lower opacity at visible wavelengths. Those super-Earths that have a constraint on the occultation depth extend to moderate equilibrium temperatures ($T_{eq,0}\sim1200$K). In this irradiation regime, particles could condense in the atmosphere, resulting in the possible formation of reflective clouds \citep{Schaefer:2009}. Constraining the dependence of brightness temperature excesses with stellar incident flux using a larger sample of objects will be key to constrain atmospheric chemistry and dynamics on super-Earths.

\subsubsection{Surfaces}
It has been shown for Kepler-10b \citep{Batalha:2011} that a lava-ocean model yields very low incident radiation absorption and strong back-scattering by molten lava, resulting in a large albedo $A_g\sim0.4$ \citep{Rouan:2011}. I find similarly large albedos for a subset of the super-Earths studied here. Still considering Kepler-10b, a wide range of solid surface compositions \citep{Hu:2012} could match the observed excess of brightness temperature  in the {\it Kepler} bandpass. Determining how mineral reflective properties evolve through molten phases with increasing temperature will be paramount to lift this degeneracy.

Those super-Earths that have large geometric albedos represent promising targets for their characterization using future facilities operating in the visible such as CHEOPS \citep{Broeg:2013} and E-ELT/HIRES \citep{Maiolino:2013} using spectroscopic direct detection \citep{Martins:2013}. New occultation follow-up studies of a well defined sample of strongly irradiated super-Earths will have the ability to constrain the range of plausible surface compositions of these exoplanets by improving the precision on planetary emission at visible and infrared wavelengths. Treating a population of planets as a whole to search for common trends may turn out to be a complementary approach to firmly constrain the surface/atmosphere properties of the ubiquitous small exoplanets.

\acknowledgments
The author thanks Julien de Wit, Farhan Feroz, Michael Gillon, Kevin Heng, Renyu Hu, Nikku Madhusudhan, Leslie Rogers, Roberto Sanchis-Ojeda, Sara Seager, Vlada Stamenkovic and Andras Zsom for providing comments that improved the manuscript. The author also thanks the anonymous referee for a very helpful review. The author gratefully acknowledges the {\it Kepler} team for generating and vetting the {\it Kepler} datasets. This research made use of the IDL Astronomical Library and the IDL-Coyote Graphics Library.

{\it Facility:} \facility{{\it Kepler}}

%\bibliography{apj-jour,se-albedos}

\end{document}